\documentstyle[wscrew,epsfig]{article}
\begin{document}

\bibliographystyle{unsrt}   % for BibTeX - sorted numerical labels

\def\ap#1#2#3   {{\em Ann. Phys. (NY)} {\bf#1} (#2) #3}
\def\apj#1#2#3  {{\em Astrophys. J.} {\bf#1} (#2) #3}
\def\apjl#1#2#3 {{\em Astrophys. J. Lett.} {\bf#1} (#2) #3}
\def\app#1#2#3  {{\em Acta. Phys. Pol.} {\bf#1} (#2) #3.}
\def\ar#1#2#3   {{\em Ann. Rev. Nucl. Part. Sci.} {\bf#1} (#2) #3}
\def\cpc#1#2#3  {{\em Computer Phys. Comm.} {\bf#1} (#2) #3}
\def\err#1#2#3  {{\it Erratum} {\bf#1} (#2) #3}
\def\ib#1#2#3   {{\it ibid.} {\bf#1} (#2) #3}
\def\jmp#1#2#3  {{\em J. Math. Phys.} {\bf#1} (#2) #3}
\def\ijmp#1#2#3 {{\em Int. J. Mod. Phys.} {\bf#1} (#2) #3}
\def\jetp#1#2#3 {{\em JETP Lett.} {\bf#1} (#2) #3}
\def\jpg#1#2#3  {{\em J. Phys. G.} {\bf#1} (#2) #3}
\def\mpl#1#2#3  {{\em Mod. Phys. Lett.} {\bf#1} (#2) #3}
\def\nat#1#2#3  {{\em Nature (London)} {\bf#1} (#2) #3}
\def\nc#1#2#3   {{\em Nuovo Cim.} {\bf#1} (#2) #3}
\def\nim#1#2#3  {{\em Nucl. Instr. Meth.} {\bf#1} (#2) #3}
\def\np#1#2#3   {{\em Nucl. Phys.} {\bf#1} (#2) #3}
\def\pcps#1#2#3 {{\em Proc. Cam. Phil. Soc.} {\bf#1} (#2) #3}
\def\pl#1#2#3   {{\em Phys. Lett.} {\bf#1} (#2) #3}
\def\prep#1#2#3 {{\em Phys. Rep.} {\bf#1} (#2) #3}
\def\prev#1#2#3 {{\em Phys. Rev.} {\bf#1} (#2) #3}
\def\prl#1#2#3  {{\em Phys. Rev. Lett.} {\bf#1} (#2) #3}
\def\prs#1#2#3  {{\em Proc. Roy. Soc.} {\bf#1} (#2) #3}
\def\ptp#1#2#3  {{\em Prog. Th. Phys.} {\bf#1} (#2) #3}
\def\ps#1#2#3   {{\em Physica Scripta} {\bf#1} (#2) #3}
\def\rmp#1#2#3  {{\em Rev. Mod. Phys.} {\bf#1} (#2) #3}
\def\rpp#1#2#3  {{\em Rep. Prog. Phys.} {\bf#1} (#2) #3}
\def\sjnp#1#2#3 {{\em Sov. J. Nucl. Phys.} {\bf#1} (#2) #3}
\def\spj#1#2#3  {{\em Sov. Phys. JEPT} {\bf#1} (#2) #3}
\def\spu#1#2#3  {{\em Sov. Phys.-Usp.} {\bf#1} (#2) #3}
\def\zp#1#2#3   {{\em Zeit. Phys.} {\bf#1} (#2) #3}

\setcounter{secnumdepth}{2} % Number sections and subsections

%%%%%%%%%%%%%%%%%%%%%%%%%%%%%%%%%%%%%%%%%%%%%%%%%%
%                                                %
%    BEGINNING OF TEXT                           %
%                                                %
%%%%%%%%%%%%%%%%%%%%%%%%%%%%%%%%%%%%%%%%%%%%%%%%%%
   
\title{\begin{flushright} HUTP-96/A-001 \end{flushright}
      \vskip.3in
      Rare, Leptonic, and Hadronic Decays of Charmed Mesons 
      \\    - A Review -}

\firstauthors{Hitoshi Yamamoto}

\firstaddress{Haravard University,
       42 Oxford St., Cambridge, Massachusetts 02138, U.S.A.}

\twocolumn[\maketitle]

\vskip.2in
\begin{center}{\bf Abstract}\end{center}

We review the physics of rare, leptonic, and hadronic decays of charmed
mesons based on the results submitted to this conference (EPS 95,
Brussels).

\section{Rare and leptonic decays}

\subsection{$D_S^+\to\mu^+\nu$}

The decay $D_S^+\to\l^+\nu$ ($l$ is $e$, $\mu$, or $\tau$)
occurs through annihilation of
constituent quarks and the rate is given by
\[
   \Gamma = |V_{cs}|^2 {G_F^2 M_{D_S} f_{D_S}^2 \over 8\pi}
            m_l^2 \Big( 1 - {m_l^2\over M_{D_S}^2} \Big)^2;
\]
thus, it is sensitive to the $D_S$ decay constant 
$f_{D_S}$ which in turn
is related to the overlap of $c$ and $\bar s$ quarks in the meson 
$\Psi(0)$ (in non-relativistic quark model) by\cite{KaplanKuhn}
\begin{equation}
    f_{D_S} = \sqrt{12\over M_{D_S}} | \Psi(0) | . 
     \label{eq:fDs}
\end{equation}
This mode was first observed by WA75\cite{WA75} using emulsion exposed
to $\pi^-$ beam from CERN SPS, followed by CLEO\cite{CLEO:Dsmunu}
and BES\cite{BES:Dsmunu}. For this conference, CLEO has updated the
measurement with higher statistics.
CLEO observes this decay mode in the decay chain
$D_S^{*+}\to D_S^+ \gamma$, $D_S^+ \to \mu^+\nu$. The hermiticity of
the detector was used to obtain the missing momentum in the 
hemisphere of the decay, which was then interpreted as
the momentum of the neutrino. A clear peak in the mass difference
$M_{Ds^*}-M_{D_S}$ is seen and the new result is shown in 
Table~\ref{tb:fDs}
together with previous results.
\begin{table}[b]
\begin{center}
\begin{tabular}{|c|c|c|}
\hline
\hline
           & $f_{D_S}$ (MeV)  & Method \\ 
\hline
  WA75\cite{WA75}
           & $232\pm45\pm20\pm48$      & $\pi^-$ on emulsion    \\ 
  CLEO(93)\cite{CLEO:Dsmunu}
           & $344\pm37\pm52\pm42$      & $e^+e^-$ (10 GeV c.m.) \\ 
  BES\cite{BES:Dsmunu}
           & $430^{+150}_{-130}\pm40$  & $e^+e^-$ (4 GeV c.m.)  \\
  CLEO(95)\cite{CLEO:Dsmunu2}
           & $284\pm30\pm30\pm16$      & $e^+e^-$ (10 GeV c.m.) \\
\hline
\end{tabular}
\end{center}
\caption{Mesurements of the decay constant $f_{D_S}$
from the decay mode $D_S^+\to\mu^+\nu$. The first error is
statistical, second systematic, the third when given is uncertainty in
$D_S$ production rate (WA75) or $B(D_S\to\phi\pi$) (CLEO).\hfil}
\label{tb:fDs}
\end{table}
The average of WA75, BES, and CLEO(95) gives 
(using both statistical and systematic errors)
\begin{equation}
  f_{D_S} ({\rm MeV})
     = (273\pm30)\sqrt{B(D_S\to\phi\pi)\over 3.5\%} .
  \label{eq:fDsmunu}
\end{equation}
The parameter $f_{D_S}$ can also be obtained
by comparing $B\to D_S^{(*)}D^*$ to semileptonic decay $B\to D^* l\nu$
assuming factorization. The new result from CLEO is 
(assuming $f_{D_S} = f_{D_S^*}$)\cite{CLEO:fDs}
\begin{equation}
  f_{D_S^{(*)}} ({\rm MeV}) 
     = (281\pm39)\sqrt{3.5\%\over B(D_S\to\phi\pi)} .
    \label{eq:fDsB}
\end{equation}
The uncertainty in the model-dependent value 
$B(D_S\to\phi\pi) = 3.5\pm0.4\%$\cite{PDG} has
been a issue for some time, but the different dependence
on $B(D_S\to\phi\pi)$ of the two $f_{D_S}$ measurements 
(\ref{eq:fDsmunu}) and (\ref{eq:fDsB}) above  
allows us to determine both  $f_{D_S}$ and 
$B(D_S\to\phi\pi)$ just by the consistency\cite{MuheimStone}
(still depends on the factorization assumption in $B\to D_S^{(*)}D^*$):
\[
  f_{D_S} = 277\pm25 ({\rm MeV}),\quad
  B(D_S\to\phi\pi) = 3.60\pm0.64\% .
\]

\subsection{Isospin violating decay $D_S^{*+}\to D_S^+\pi^0$}

The decay $D_S^{*+}\to D_S^+\pi^0$ is prohibited by isospin. 
The amplitude of the dominant channel $D_S^{*+}\to D_S^+\gamma$
is proportional to the total effective magnetic moment
which is naively given by
\[
  \vec\mu = {q_c\over m_c} \vec s_c + {q_s\over m_s} \vec s_s
\]
where $q_c (=2/3e)$ and $q_s (=-1/3e)$ are the charges of
$c$ and $s$ quarks, and $\vec s$ are the corresponding
spins which are aligned in the case of $D_S^*$. 
Using constituent masses $m_c = 1.5$ GeV and $m_s = 0.4$ GeV, 
one can see
that the dominant mode is highly suppressed. Such near cancellation in
the dominant mode makes it difficult to predict the branching fraction
for $D_S^{*+}\to D_S^+\pi^0$.
One estimate\cite{ChoWise} based on $\eta$-$\pi^0$ mixing (together with
HQET and chiral perturbation theory) gives a range of a few \%.
CLEO\cite{CLEO:DSpi} has searched for this mode using 
the decay mode $D_S^+\to\phi\pi^+$,
and looks at the mass difference $M(D_S^{*+}) - M(D_S^+)$. A clean signal
with $14^{+4.6}_{-4.0}$ events was observed. It is then normalized to the
dominant radiative decay to give
\[
  \Gamma(D_S^{*+}\to D_S^+\pi^0)/\Gamma(D_S^{*+}\to D_S^+\gamma)
   = 0.062^{+0.020}_{-0.018}\pm0.022 .
\]
The transition $D_S^*\to D_S \pi$ tells us that the
parity of $D_S^*$ is $(-)^J$ where $J$ is the spin of $D_S^*$. Also,
the decay $D_S^{*+}\to D_S^+\gamma$ indicates that $J\geq1$. Thus, 
possible spin/parity of $D_S^*$ is $1^-, 2^+, 3^-$\ldots, with $1^-$
being the most natural candidate. This is also consistent with the quark
model prediction for the mass splitting\cite{FrankODonnell}
\[
    M(^3 S_1)-M(^1S_0)={32\pi\alpha_s\over9m_c m_s}|\Psi(0)|^2 .
\]
With $\alpha_s = 0.5$ and eq. (\ref{eq:fDs}), this gives 0.13 GeV 
which is quite consistent with the experimental value of
$0.1416\pm0.0018$ GeV.

\begin{table*}
\begin{center}
\begin{tabular}{|c|c|c|c|}
\hline
\hline
   $R$ (def.) & Exp. & $R$(\%) & 
           $R$ in unit of $\lambda^4$ or $\lambda^2$\\ 
\hline
  $\Gamma(D^0\to K^+\pi^-) / \Gamma(D^0\to K^-\pi^+)$ 
           & CLEO\cite{CLEO:DCSD}
           & $0.77\pm0.25\pm0.25$      
           & $(2.92\pm0.95\pm0.95)\lambda^4$   \\ 
  $\Gamma(D^+\to K^+\pi^-\pi^+) / \Gamma(D^+\to K^-\pi^+\pi^+)$ 
           & $\begin{array}{c} \hbox{E687\cite{E687:DCSD}} \\ 
                               \hbox{E791\cite{E791:DCSD}} \end{array}$
           & $\begin{array}{c} 0.72\pm0.23\pm0.17 \\
                               1.03\pm0.24\pm0.13
              \end{array}$ 
           & $\begin{array}{c} (2.73\pm0.87\pm0.64)\lambda^4 \\
                               (3.9\pm0.9\pm0.5)\lambda^4
              \end{array}$ \\
  $\Gamma(D^+\to K^{*0}\pi^+)/\Gamma(D^+\to K^-\pi^+\pi^+)$ 
           & E687\cite{E687:DCSD}
           & $<0.021$      &                   \\
  $\Gamma(D^+\to K^+\rho^0)/\Gamma(D^+\to K^-\pi^+\pi^+)$ 
           & E687\cite{E687:DCSD}
           & $<0.067$      &                   \\
  $\Gamma(D^+\to K^+K^-K^+) / \Gamma(D^+\to\phi\pi^+)$ 
           & E687\cite{E687:3K}
           & $<2.5$      &                     \\
  $\Gamma(D^+\to \phi K^+) / \Gamma(D^+\to\phi\pi^+)$ 
           & E687\cite{E687:3K}
           & $<2.1$      & $<0.41 \lambda^2$   \\
  $\Gamma(D^+\to \phi K^+) / \Gamma(D^+\to\phi\pi^+)$ 
           & E691\cite{E691:3K}
           & $5.8^{+3.2}_{-2.6}\pm0.7$      
           & $(1.1^{+0.6}_{-0.5}\pm0.1) \lambda^2$   \\
\hline
  $\Gamma(D_S^+\to K^+\pi^-\pi^+) / \Gamma(D_S^+\to\phi\pi^+)$ 
           & E687\cite{E687:DCSD}
           & $28\pm6\pm5$      &               \\
  $\Gamma(D_S^+\to\bar K^{*0}\pi^+) / \Gamma(D_S^+\to\phi\pi^+)$ 
           & E687\cite{E687:DCSD}
           & $18\pm5\pm4$      & $(3.5\pm1.0\pm0.8)\lambda^2$   \\
  $\Gamma(D_S^+\to K^+\rho^0) / \Gamma(D_S^+\to\phi\pi^+)$ 
           & E687\cite{E687:DCSD}
           & $<8$      &               \\
  $\Gamma(D_S^+\to K^+K^-K^+) / \Gamma(D_S^+\to\phi\pi^+)$ 
           & E687\cite{E687:3K}
           & $<1.6$      &                     \\
  $\Gamma(D_S^+\to \phi K^+) / \Gamma(D_S^+\to\phi\pi^+)$ 
           & E687\cite{E687:3K}
           & $<1.3$      & $<0.25 \lambda^2$   \\
\hline
\end{tabular}
\end{center}
\caption{Mesurements of DCSD modes of $D^+$ and 
singly-Cabibbo-suppressed modes of $D_S^+$. 
The Cabibbo suppression factor
used is $\lambda\equiv\tan\theta_C=0.226$. Also listed are two 
Cabibbo-suppressed modes of $D_S^+$. E687, E691 (photo-production) and E791
(hadro-production) are fixed-target experiments at Fermilab.}
\label{tb:DCSD}
\end{table*}

\begin{figure}[h]
\centering
\mbox{\psfig{figure=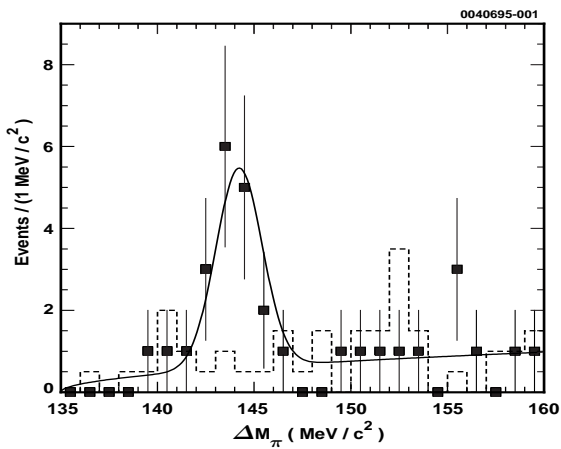,height=2.5in,width=3.5in}}
  \caption{Mass difference $\Delta M_\pi \equiv M(D_S^{*+})
   - M(D_S^+)$ for the isospin violating decay $D_S^{*+}\to
   D_S^+\pi^0$ (CLEO). The dotted line is
   the background estimated from the $D_S^+$ and $\pi^0$ sidebands.}
  \label{fg:Dspi0}
\end{figure}

\section{Hadronic decays}

In the factorization model of Bauer, Stech, and Wirbel (BSW)\cite{BSW}, 
the hadronic two-body decays of charmed mesons occur through
the effective Hamiltonian given by
\[
   {\cal H}_{\rm had} = {G_F \over \sqrt2} V^*_{ud}V_{cs} 
   [ a_1(\overline d u)_{\rm had} (\overline c b)_{\rm had} +  
     a_2(\overline d b)_{\rm had} (\overline c u)_{\rm had} ]
   \nonumber
\]
where $(..)_{\rm had}$ indicates the factorized current ready to form a
final-state meson. The $\bar d$ can be changed to $\bar s$ and
$s$ may be changed to $d$ with appropriate change in the CKM elements.
It is convenient to classify the two-body decays to three categories:
(Class-1) The amplitude is $f_1 a_1$; e.g. 
                  $D^0 \to K^-\pi^+$.
(Class-2) The amplitude is $f_2 a_2$; e.g.
                  $D^0 \to \bar K^0 \pi^0$ (`color-suppressed').
(Class-3) The amplitude is ($f_1 a_1 + f_2 a_2$); e.g.
                  $D^+ \to \bar K^0 \pi^+$.
where the constants $f_i$ depend on form factors and decay constants and
typically well within factor of two of each other if related by flavor SU(3)
(apart from isospin factors).
Fitting the recent measurements of $D^0\to K^-\pi^+, \bar K^0\pi^0$, and
$D^+\to \bar K^0 \pi^+$, the values of $a_1, a_2$ are found to be
\[
a_1 \sim 1.15,\quad  a_2 \sim -0.51
\]
which indicates that the Class 3 decays involve destructive interferences.
The color suppression and interference suppression each amounts to
very roughly a factor of 1/4 in decay rate.
The parameters $a_1$ and $a_2$, however, 
in principle depend on final state
and may not be the same for other decay modes such as $D\to KK, \pi\pi$.
Another factor that can affect the rate is the phase space; in the comparisons
that follow, however, the phase-space factors are mostly within 20\%\
including P-wave decays.

\begin{figure}[h]
\centering
\mbox{\psfig{figure=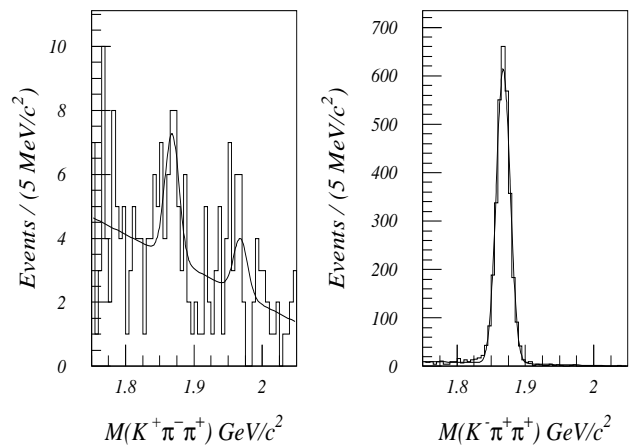,height=2.5in,width=3.5in}}
  \caption{Mass distribution for the DCSD mode 
   $D^+\to K^+\pi^-\pi^+$ (left) and the corresponding Cabibbo-allowed
   decay (right) (E687). The enhancement at 1.97 GeV is
   $D_S^+\to K^+\pi^-\pi^+$. Cuts are optimized for the $D^+$
   decay.}
  \label{fg:DCSD}
\end{figure}

\subsection{Singly and Doubly-Cabibbo-Suppressed Decays}

\begin{figure}
\centering
\mbox{\psfig{figure=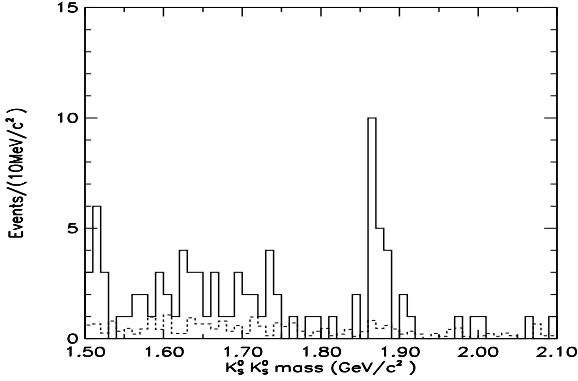,height=2.2in,width=3.5in}}
  \caption{The decay $D^0\to K_SK_S$ tagged by $D^{*+}\to D^0\pi^+$
    and $D^{*0}\to D^0\pi^0$ (CLEO). The dotted line shows the
    $D^*$ side band.}
  \label{fg:KSKS}
\end{figure}

Possibly three DCSD (Doubly Cabibbo Suppressed Decay) 
modes have been observed
thus far: $D^0\to K^+\pi^-$, $D^+\to K^+\pi^-\pi^+$, and 
$D^+\to \phi K^+$.  
The results are given in Table~\ref{tb:DCSD}.
It should be noted that the decay $D^0\to K^+\pi^-$ can also occur
by $D^0-\bar D^0$ mixing. Including the interference, the time-dependent
and time-integrated decay rates are given by
\begin{equation}
 \Gamma(D^0\to K^+\pi^-)(t) = 
         |a|^2 \Big| \,\rho_D + {1\over\sqrt2}\rho_M\, t\,
               \Big|^2 e^{-t},
     \label{eq:DCSDmix}
\end{equation}
\[
 R \equiv { B(D^0\to K^+\pi^-) \over B(D^0\to K^-\pi^+) }
    = r_D + \sqrt{2 r_D r_M} \cos(\phi_D - \phi_M) + r_M ,
    \nonumber
\]
where 
\[
\rho_D \equiv {a_{D}\over a} \equiv\sqrt{r_D}e^{i\phi_D},\;
\rho_M \equiv {\delta\gamma/2 + i\delta m \over \sqrt2\gamma}
   \equiv \sqrt{r_M}e^{i\phi_M},
\]
with $a\equiv Amp(D^0\to K^-\pi^+)$, 
$a_{D}\equiv Amp(D^0\to K^+\pi^-)$, $\gamma$ is the average $D^0$
decay rate, $t$ is in unit of $1/\gamma$, and assumed $r_{M,D}\ll 1$.
The expressions above are identical for $\bar D^0\to K^+\pi^-$ with
the redefinitions $a\equiv Amp(\bar D^0\to K^+\pi^-)$
$a_{D}\equiv Amp(\bar D^0\to K^-\pi^+)$. Assuming CP symmetry,
values of $\rho_{D,M}$ are the same for the charge conjugate
cases.
Note that the parameter $r_M$ is the standard mixing parameter
$r_M = (D^0 \hbox{ decays as } \bar D^0) /
       (D^0 \hbox{ decays as } D^0)$.
E691\cite{E691}
gives $r_M<0.39\%$ neglecting the interference term. The limit,
however, increases about factor of three when no constraint is imposed
on the interference phase. 

Naively, one expects that a DCSD mode to be suppressed by a factor
$\lambda^4$ ($\lambda\equiv\tan\theta_c$) relative to the corresponding
Cabibbo-favored mode. Assuming no $D^0-\bar D^0$ mixing, the DCSD
decay $D^0\to K^+\pi^-$ is about 3 times larger than the naive
expectation even though statistically not very significant. 
A straight application of the BSW model\cite{BSW}
predicts 1.5 $\lambda^4$ where
the enhancement is almost entirely accounted for by the difference
in the decay constants: $(f_K/f_\pi)^2$. 
The measured DCSD enhancement for
$D^+\to K^+\pi^-\pi^+$ is slightly more significant. 
This is expected
since the only resonance submode in the Cabibbo-favored mode is
$D^+\to "\bar K^{*0}"\pi^+$ ($"\bar K^{*0}"$ stands for any excited state of 
$\bar K^0$) which is Class-3 and suppressed by the 
destructive interference. In fact, the $D^+\to K^-\pi^+\pi^+$ is 
the only mode among $D\to K\pi\pi$ 3-body decays that is not dominated
by resonance submodes\cite{E687:KppDalitz}.

The decay of $D^+$ to 3 charged kaons is a DCSD by charge conservation.
At quark diagram level, the spectator DCSD has a pair of
$d\bar d$ quarks while the 3$K^\pm$ final state
has no valence $d\bar d$ quarks. Thus, the strong final state
interaction (FSI) should annihilate the $d\bar d$ pair and 
create $s \bar s$ pair. Another possibility is the annihilation
mode $c\bar d\to u\bar s$ followed by creation of a $s\bar s$ pairs, 
but this mode itself is doubly-Cabibbo-suppressed
and expected to be small. This leads to suppression of the
3$K^\pm$ DCSD mode relative to the naive expectation. As can be seen
in the table, the experimental upper limit for 
$B(D^+\to\phi K^+)/B(D^+\to\phi\pi^+)\;(\phi\to K^+K^-)$ 
is less than half (E687) or about equal to (E691) the naive value.
Note that the normalization mode $\phi\pi^+$
itself is color-suppressed (Class-2).
The abnormally large value
$B(D^+\to K^+K^-K^+)/B(D^+\to K^-\pi^+\pi^+)=(5.7\pm2.0\pm0.7)\%$
by WA82\cite{WA82:DCSD} (which is about 20 times
larger than the naive expectation) is not consistent with these results. 
Also listed is the singly-Cabibbo-suppressed modes of  
$D_S^+$ which have the same final states as the DCSD modes of
$D^+$ and thus can be measured in the same analyses. 
The decay $D_S^+\to\phi K^+$ is Class-3 and expected to be 
suppressed by destructive interference (there are two $\bar s$
quarks in the final state) which is consistent with
the upper limit. The decay $D_S^+\to \bar K^{*0}\pi^+$
is Class-1 and one would expect that the naive suppression factor
should work; the measurement, however, is on the high side.

In the flavor SU(3) limit, the Cabibbo-suppressed modes 
$D\to KK$ and $D\to \pi\pi$ are expected to be the same.
In comparing with the data, however, one needs to take into 
account the FSI which rotates the phases of isospin amplitudes.
For $D^0\to KK$, we have
\[
 \begin{array}{rcl}
 Amp(K^+K^-) &=& \sqrt{1\over2}(A_1 + A_0) \\
 Amp(K^0 \bar K^0) &=& \sqrt{1\over2}(A_1 - A_0),
 \end{array}
\]
where $A_I(I=0,1)$ is the isospin=I amplitude. Note that phase rotations
of $A_I$ change the decay rate of each mode, but keep the sum unchanged:
\[
\Gamma(K^+K^-) + \Gamma(K^0 \bar K^0) = {1\over2}(|A_0|^2 + |A_1|^2).
\]
The mode $D^+\to \bar K^0 K^+$ is purely $I = 1$, and thus the rate does not
change under the FSI. Here we are assuming that the FSI is elastic (i.e.
KK channels stay within KK channels), and if the FSI is inelastic the
above relations are no longer correct. The inelastic FSI, however,
is expected to be small.
The situation is similar for $D^0\to \pi\pi$. 

\begin{table}
\begin{center}
\begin{tabular}{|r|c|c|}
\hline
\hline
    Exp.   & $D^0\to K^+K^-$(\%)  & $D^0\to K^0 \bar K^0$(\%)     \\ 
\hline
    CLEO\cite{CLEO:KK}
     & $0.455\pm0.029\pm0.024$ & $0.048\pm0.012\pm0.006$ \\
    E687\cite{E687:KK} 
     & $0.437\pm0.028\pm0.060$ & $0.207\pm0.069\pm0.073$ \\
 average & $0.451\pm0.033$ & $0.051\pm0.021$ \\
\hline
\end{tabular}
\end{center}
\caption{Cabibbo-suppressed decays $D^0 \to K^+K^-, K^0\bar K^0$.}
\label{tb:KK}
\end{table}

\begin{table}
\begin{center}
\begin{tabular}{|r|c|c|}
\hline
\hline
    R(def.)   & Exp.  & ratio     \\
\hline
  ${\Gamma(D^+\to\bar \pi^0\pi^+)\over\Gamma(D^+\to\bar K^-\pi^+\pi^+)}$ &
    CLEO\cite{CLEO:pipi} &     $0.028\pm0.006\pm0.006$    \\
  ${\Gamma(D^+\to\bar K^0 K^+)\over\Gamma(D^+\to\bar K^0\pi^+)}$ &
    E687\cite{E687:K0K} &      $0.25\pm0.04\pm0.02$    \\
  ${\Gamma(D_S^+\to K^0\pi^+)\over\Gamma(D_S^+\to\bar K^0 K^+)}$ &
    E687\cite{E687:K0K} &      $0.18\pm0.21$           \\
  ${\Gamma(D_S^+\to K^0\pi^+)\over\Gamma(D_S^+\to\phi\pi^+)}$ &
    MKIII\cite{MKIII:DsK0pi} & $<0.21$                 \\
\hline
\end{tabular}
\end{center}
\caption{Cabibbo-suppressed decays of $D^+$ and $D_S^+$ to 
two pseudoscalars.}
\label{tb:K0K}
\end{table}

The $K^0\bar K^0$ mode is detected by $K_S K_S$. Since the two neutral
kaons are produced coherently in a symmetric L=0 state, one has to take into
account the quantum correlation when converting the $K_S K_S$ rate to
the $K^0\bar K^0$ rate. The state can be written as
\[
   D^0 \to K^0\bar K^0 + \bar K^0 K^0 = K_S K_S + K_L K_L
\]
which means that half the time it shows up as $K_S  K_S$ and half
the as $K_L K_L$. This is correct even with CP violation. Thus we have
\[
   B(D^0\to K^0\bar K^0) = 2 B(D^0\to K_S K_S).
\]
The decay $D^0\to K^0\bar K^0$ cannot occur by a spectator diagram
without FSI. It could occur by exchange diagrams: $c\bar u\to s\bar s$
with a $d\bar d$ pair creation from vacuum, or $c\bar u\to d\bar d$ with a
$s\bar s$ pair creation from vacuum. The two exchange amplitudes,
however, cancel exactly
in the limit of the flavor SU(3) in 2 generations (an example of 
GIM cancellation)\cite{Pham}. Thus, it is likely that 
the $K^0\bar K^0$ mode is dominated
by the feed though from the $K^+K^-$ mode due to FSI. Latest results for
$D^0\to KK$ are given in Table~\ref{tb:KK}.  Together with $D\to\pi\pi$
data from the particle data group\cite{PDG}, we get
\[
{B(K^+K^-)+B(K^0\bar K^0)\over B(\pi^+\pi^-)+B(\pi^0\pi^0)} = 2.03\pm0.27
\]
which is larger than the naive value of unity. The BSW model
prediction is 1.3 which is still smaller than the
experimental value above.

\begin{figure}
\centering
\mbox{\psfig{figure=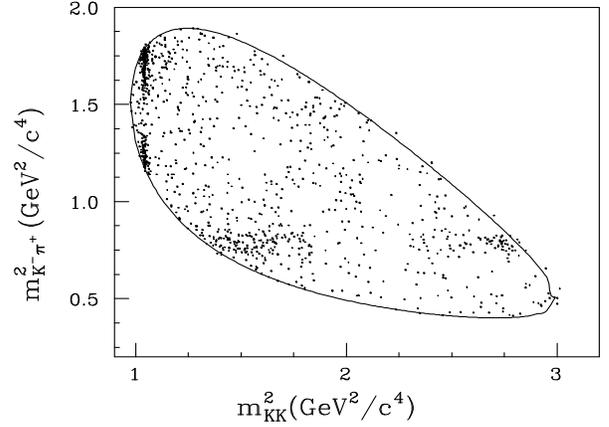,height=2.5in,width=3.5in}}
  \caption{Dalitz plot of the Cabibbo-suppressed decay 
   $D^+\to K^-K^+\pi^+$ by E687. The $\phi\pi^+$ band and the
   $\bar K^{*0}\pi^+$ band are clearly seen.}
  \label{fg:Dalitz}
\end{figure}

Cabibbo-suppressed 2-body decays of $D^+$ and $D_S^+$ are shown in 
Table~\ref{tb:K0K}. As mentioned earlier, the decays $D^+\to\pi^0\pi^+$
and $\bar K^0 K^+$ do not suffer from (elastic) FSI. 
The $D^+\to\pi^0\pi^+$ mode is Class-3, and thus expected to be
suppressed due to the destructive interference, 
while the $D^+\to\bar K^0 K^+$ mode is not. Using 
$B(D^+\to \bar K^0\pi^+)$ = $2.74\pm0.29\%$ and 
$B(D^+\to K^-\pi^+\pi^+)$ = $9.1\pm0.6\%$\cite{PDG}, the ratio of
the two Cabibbo-suppressed decays is
\[
  {\Gamma(D^+\to\bar K^0 K^+) \over 2\Gamma(D^+\to \pi^0\pi^+)}
    = 1.4\pm0.5 ,
\]
where the factor of 2 in the denominator is the isospin factor,
or equivalently due to the fact that $\pi^0$ is half $u\bar u$
and half $d\bar d$.
This factor is expected to be enhanced due to the suppression of
$\pi^0\pi^+$.
The ratio $\Gamma(D^+\to\bar K^0 K^+)/\Gamma(D^+\to\bar K^0\pi^+)$
= $0.25\pm0.04\pm0.02$ itself is larger than the naive Cabibbo
suppression of $\lambda^2\sim 0.05$. This is probably because 
the normalization
mode $D^+\to\bar K^0\pi^+$ is suppressed by the destructive
interference (Class-3). 

For $D_S^+$, the normalization mode $D_S^+\to \bar K^0 K^+$ is
color-suppressed (i.e. Class-2), while $D_S^+\to K^0\pi^+$
is not (Class-1). Thus, we expect the ratio to be enhanced with
respect to the naive factor of $\lambda^2\sim0.05$.
The experimental data is not yet conclusive at this point.
It should be noted that 
$\bar K^0$ or $K^0$ in any decay mode is 
usually measured as $K_S$ and thus 
DCSD mode could interfere with the corresponding
Cabibbo-favored mode.\cite{BigiYam} As a result, the oft-used relation
$B(\bar K^0 X) = 2 B(K_S X)$ does not hold. The correction is
typically of order 5 to 10\%, but for some channels it could
be as large as 25\%. One such example is the normalization mode 
$D_S^+\to K_S K^+$, where the 
Cabibbo-favored mode $D_S^+\to \bar K^0 K^+$ is color-suppressed
(i.e. Class-2) but the DCSD mode $D_S^+\to K^0 K^+$ is not.
The exact amount of correction depends on unknown
FSI phase, and it is difficult to reliably estimate.

For $D^+,D_S^+\to K^-K^+\pi^+$ modes, we now have Dalitz plot 
analysis from E687.\cite{E687:KKpDalitz} 
Figure~\ref{fg:Dalitz} shows the Cabibbo-suppressed
mode $D^+\to K^-K^+\pi^+$. The position of the $K^*(892)$ band was
found to be shifted in a way consistent with existence of $K^*(1430)$.
After the fit including the interferences, the ratios of partial
widths are found to be
\[
\begin{array}{rcl}
    {\Gamma(D^+\to\bar K^{*0}(892)K^+)\over\Gamma(D^+\to K^-\pi^+\pi^+)}
    & = & 0.044\pm0.003\pm0.004 \\
    {\Gamma(D^+\to\phi\pi^+)\over\Gamma(D^+\to K^-\pi^+\pi^+)}
    & = & 0.058\pm0.006\pm0.006 .
\end{array}
\]
The $D^+\to\phi\pi^+$ mode is color-suppressed, but the decay constant
of $\phi$ (which is `emitted' from the hadronic current $D\to\pi$ in
the factorization picture) is quite large ($f_\phi\sim  233$ MeV), 
could bring it up to the same level as 
$D^+\to K^{*0}K^+$ which is a Class-1 decay.

\subsection{CP Violation}

CP asymmetries in $D$ decays can occur through 1) direct CP
violation\cite{GoldenGrin,Buccella}, 2) mixing\cite{BigiSanda}, or 3)
interference of Cabibbo-favored mode and DCSD in modes involving
$K_S$\cite{BigiYam}. The asymmetries are expected to be quite
small in the standard model ($10^{-3}$ level or less); any
asymmetry much larger is thus a signal of physics beyond the
standard model.

Direct CP violation should involve at least two quark-level diagrams
with different CKM phases and different FSI phases. 
The two diagrams may be two spectators\cite{GoldenGrin}, 
spectator plus penguin\cite{Buccella}, or spectator plus 
annihilation\cite{ChauCheng}. At least in the standard model, the
asymmetries are larger for Cabibbo-suppressed modes; since the
main goal now is to search effects beyond the standard model,
however, Cabibbo-favored modes also need to be checked.
Table~\ref{tb:CP} shows recent measurements of CP asymmetries
of $D^{+,0}$ decays.
\begin{table}
\begin{center}
\begin{tabular}{|r|r|r|}
\hline
\hline
    Mode   & Exp.  & Asymmetry                               \\ 
\hline
   $D^+\to
    \begin{array}{c}
       K^+K^-\pi^+        \\  \bar K^{*0}K^+ \\ \phi\pi^+
    \end{array}$    & E687\cite{E687:CP}\quad &
   $\begin{array}{r}
       -0.031\pm0.068 \\ -0.12\pm0.13 \\ 0.066\pm0.086
    \end{array}$                                             \\
\hline
   $K^+K^-$ & 
   $\begin{array}{r}
      \hbox{E687\cite{E687:CP}} \\ \hbox{CLEO\cite{CLEO:CP}}
    \end{array}$ &
   $\begin{array}{r}
      0.024\pm0.084             \\  0.069\pm0.059
    \end{array}$                                             \\
   $\begin{array}{c} D^0\to \\ \quad \end{array}$
   $\begin{array}{c}
       K_S\phi   \\ K_S \pi^0 \\ K^-\pi^+
    \end{array}$  &
   $\begin{array}{r}
       \hbox{CLEO\cite{CLEO:CP}} \\
       \hbox{CLEO\cite{CLEO:CP}} \\
       \hbox{CLEO\cite{CLEO:CP}} 
    \end{array}$  &
   $\begin{array}{r}
       -0.007\pm0.090 \\ -0.013\pm0.030 \\ 0.009\pm0.011
    \end{array}$                                             \\
\hline
\end{tabular}
\end{center}
\caption{CP asymmetries of $D^{0,+}$ decays. The asymmetries by
  E687 are normalized to $D^+\to K^-\pi^+\pi^+$ and $D^0\to K^-\pi^+$.
  CLEO tags the flavor of $D$ by the decay $D^{*+}\to D^0\pi^+$.}
\label{tb:CP}
\end{table}
Since there is an asymmetry in photo-production of $D$ and $\bar D$ in
E687 numbers are normalized to corresponding Cabibbo-favored decays
$D^+\to K^-\pi^+\pi^+$ and $D^0\to K^-\pi^+$. All numbers are consistent
with zero. 

For $D^0\to K\pi$, CP violation in mixing introduces additional
factor between $\rho_D$ and $\rho_M$ in (\ref{eq:DCSDmix})
which differ for $D^0\to K^+\pi^-$ and $\bar D^0\to K^-\pi^+$
(assuming that there is no direct CP violation):
\[
    \rho_M \to {q\over p}\rho_M \quad (D^0),  \qquad
    \rho_M \to {p\over q}\rho_M \quad (\bar D^0)
\]
where $p$ and $q$ are the coefficients of $D^0$ and $\bar D^0$ for the
mass eigenstates $D_{H,L}$:
\[
     D_H = pD^0 + q\bar D^0,\quad D_L = pD^0 - q\bar D^0 .
\]
If $p\ne q$, then the additional factor is in general different 
for $D^0$ and $\bar D^0$ resulting in
CP asymmetry in the time-dependent 
$\Gamma(D^0\to K^+\pi^-)(t)$ vs. its CP conjugate
channel as well as in the time-integrated rates. 
In the limit of $\Im(a_D/a)=0$ and $\delta\gamma=0$,
there is no term linear in $t$ in the expression (\ref{eq:DCSDmix}). 
Even then, 
CP violation in mixing would give rise to a term linear in $t$
which is also $CP$ odd. Such asymmetry that increases
linearly with time would 
indicate both CP violation and mixing simultaneously\cite{Wolfen}.
The table also lists asymmetries for CP self-conjugate final
states; the results are all consistent with zero.

\section{Summary}

The charm physics has matured. We now see rare decays such as
pure leptonic decay and isospin violating decay. A few 
Doubly-Cabibbo-Suppressed decays have been observed, and
detailed studies of singly-Cabibbo-suppressed modes have been
done and ongoing. Most of the modes are qualitatively understandable
in terms of the color-suppression and the destructive interference
together with known SU(3) breaking effects such as the differences in
decay constants.
The experimental sensitivities for
CP asymmetries are still a few orders of magnitude away from
the level expected by the standard model.
\vskip.3in
{\it Acknowledgements}: The author would like to thank 
I. Bigi, S. P. Ratti, F. Muheim and V. Jain
for useful and stimulating conversations. This work was partly supported by
the U.S. department of energy.

\vskip 0.1in
\center

\end{document}